\documentclass[twocolumn,aps,prl,amsmath,amssymb,showpacs]{revtex4}
\usepackage{color}
\usepackage{mathrsfs}
\usepackage{amsmath}
\usepackage{amssymb}
\usepackage{graphicx}
\usepackage{esint}
\usepackage[unicode=true,pdfusetitle,
 bookmarks=true,bookmarksnumbered=false,bookmarksopen=false,
 breaklinks=false,pdfborder={0 0 0},pdfborderstyle={},backref=false,colorlinks=true]
 {hyperref}
\hypersetup{
 colorlinks,linkcolor=red,citecolor=blue}

\makeatletter

\newcommand*\LyXThinSpace{\,\hspace{0pt}}

\usepackage{epsfig}\usepackage{amsfonts}

\makeatother

\begin{document}

\title{Phase-Controlled Phonon Laser}

\author{Yan-Lei Zhang, $^{1,2}$ }

\author{Chang-Ling Zou, $^{1,2,3}$ }

\author{Chuan-Sheng Yang, $^{1,2}$ }

\author{Hui Jing, $^{4}$ }
\email{jinghui73@gmail.com}

\author{Chun-Hua Dong $^{1,2}$ }

\author{Guang-Can Guo $^{1,2}$ }

\author{Xu-Bo Zou, $^{1,2}$ }
\email{xbz@ustc.edu.cn}

\affiliation{$^{1}$ Key Laboratory of Quantum Information, University of Science
and Technology of China, Hefei 230026, People\textquoteright s Republic
of China}

\affiliation{$^{2}$ Synergetic Innovation Center of Quantum Information and Quantum
Physics, University of Science and Technology of China, Hefei, Anhui
230026, China}

\affiliation{$^{3}$ Department of Applied Physics, Yale University, New Haven,
Connecticut 06511, USA}

\affiliation{$^{4}$ Key Laboratory of Low-Dimensional Quantum Structures and
Quantum Control of Ministry of Education, Department of Physics and
Synergetic Innovation Center for Quantum Effects and Applications,
Hunan Normal University, Changsha 410081, China}

\date{\today}
\begin{abstract}
A phase-controlled ultralow-threshold phonon laser is proposed by
using tunable optical amplifiers in coupled-cavity-optomechanical
system. Giant enhancement of coherent photon-phonon interactions is
achieved by engineering the strengths and phases of external parametric
driving. This in turn enables single-photon optomechanics and low-power
phonon lasing, opening up novel prospects for applications, e.g. quantum
phononics and ultrasensitive motion detection.
\end{abstract}


\maketitle
\emph{Introduction.}- As a promising platform to study fascinating
macroscopic quantum phenomena \cite{Chen2013}, cavity optomechanics
\cite{Schwab2012,Vahala2008} has received tremendous attentions in
recent years. All kinds of optomechanical couplings and applications
\cite{Marquardt2014} have been opened up due to remarkable experimental
advances in e.g., mechanical ground-state cooling \cite{Teufel2011,Wang2009},
optomechanical non-reciprocity \cite{Dong2015,Shen2016,Kim2015},
optomechanically induced transparency \cite{Weis2010,Safavi-Naeini2011},
nonclassical state preparation \cite{Purdy2013,Naeini2013}, coherent
state transfer between light and sound \cite{Fiore2011,Zhou2013},
and various phonon-mediated hybrid devices \cite{Dong2012}. To extend
more applications, on the one hand, the unique regime of single-photon
quantum optomechanics \cite{Rabl2011,Nunnenkamp2011}, however, is
still pursued in current experimental efforts; on the other hand,
we need to realize convenient tuning, especially the switching between
different optomechanical couplings.

To realize single-photon coupling, many theoretical schemes have been
proposed based on, for examples, undriven two-cavity set-ups \cite{Bhattacharya2008},
optomechanical arrays \cite{Xuereb2012}, Josephson effect \cite{Heikkil=0000E42014},
and the transient scheme \cite{Xu2015}. Very recently, the parametric
drive has been used to enhance the nonlinear coupling \cite{Xin2015,Clerk2016,Li2016}
in optomechanical systems. By exploiting coupled cavities, many applications
have been studied, such as single-photon generation \cite{Bamba99},
steady-state entanglement \cite{shen2011}, thermal phonon squeezing
\cite{Mahboob2014}, and phonon laser \cite{He2016}. In the study
of phononic devices \cite{Li2012,LaHaye2009,Hatanaka2013}, the phonon
laser \cite{Vahala2009,Khurgin2012} plays a key role in integrating
coherent phonon sources, detectors, and waveguides \cite{Eichenfeld2009}.
Phonon lasing have been demonstrated in the electromechanical resonator
\cite{Mahboob2013}, the nanomechanical resonator \cite{Cohen2015},
the vertical cavity structure \cite{Maryam2013}, and the compound
microcavity system \cite{Grudinin2010}, and some schemes are also
proposed to realize phonon laser in the quantum-dot system \cite{Khaetskii2013,Kabuss2012}.
In particular, the ultralow-threshold phonon laser \cite{Jing2014,Wang2017}
still gives rise to the broad interest and remains largely unplored.

In this paper, we present a scheme for both enhancing optomechanical
couplings into the single-photon strong-coupling regime and realizing
the switching between different optomechanical interactions using
optical parametric amplifiers (OPAs). The key idea is to put two OPAs
into both the auxiliary cavity and the optomechanical system, which
leads to the squeezing of transformational optical modes. Due to the
squeezing, we can obtain exponentially enhanced radiation-pressure,
parametric amplification, and three-mode optomechanical couplings,
which are controlled by the phase difference from the two OPAs. As
one of applications, we study a phase-controlled ultralow-threshold
phonon laser in detail. In addition, we consider the noise of the
squeezed modes, which can be suppressed greatly via dissipative squeezing
or an additional optical mode. With current experimentally accessible
parameters, our scheme should be feasible to study quantum optomechanics.

\begin{figure}
\includegraphics[width=0.9\columnwidth]{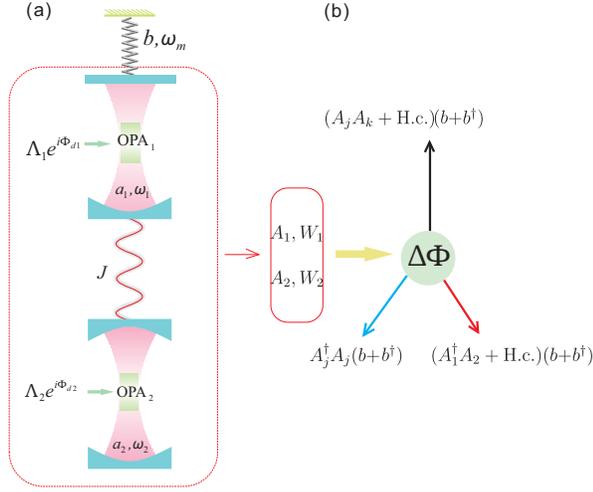}

\caption{(Color online) (a) Schematic diagram of the optomechanical system
with coupled cavities. Each cavity contains an OPA with driving amplitude
$\varLambda_{j}$, frequency $\omega_{dj}$, and phase $\Phi_{dj}$,
respectively. The photon-hopping interaction $J$ leads to the supermodes
$A_{j}$ with the frequency $W_{j}$. (b) The phase difference $\Delta\Phi=\Phi_{d1}-\Phi_{d2}$
controls the radiation-pressure $A_{j}^{\dagger}A_{j}\left(b^{\dagger}+b\right)$,
parametric amplification $\left(A_{j}A_{k}+\mathrm{H.c.}\right)\left(b^{\dagger}+b\right)$,
and three-mode $\left(A_{1}^{\dagger}A_{2}+\mathrm{H.c.}\right)\left(b^{\dagger}+b\right)$
optomechanical couplings.}

\label{Fig1}
\end{figure}

\emph{Model.-} We consider an optomechanical system with two coupled
cavities, and each cavity contains a driven nonlinear optical medium
for OPA, as shown in Fig.$\,$\ref{Fig1}(a), which can be described
by the Hamiltonian ($\hbar=1$)

\begin{align}
H= & H_{c}+H_{m}+J\left(a_{1}a_{2}^{\dagger}+a_{1}^{\dagger}a_{2}\right),\\
H_{c}= & \sum_{j}\omega_{j}a_{j}^{\dagger}a_{j}+\varLambda_{j}\left(a_{j}^{\dagger2}e^{-i\Phi_{dj}-i\omega_{dj}t}+\mathrm{H.c.}\right),\\
H_{m}= & \omega_{m}b^{\dagger}b-g_{0}a_{2}^{\dagger}a_{2}\left(b^{\dagger}+b\right),
\end{align}
where $a_{j}$ and $b$ are the annihilation operators for the $j$th
($j=1,~2$) cavity mode with frequency $\omega_{j}$ and the mechanical
mode with frequency $\omega_{m}$, respectively, and $J$ is the photon-hopping
interaction strength between two cavities. $H_{c}$ describes the
optical modes containing two different OPAs. $H_{m}$ describes the
optomechanical system associated with the 2nd cavity, in which $g_{0}$
is the radiation-pressure optomechanical coupling strength.

\emph{Phase-controlled optomechanical systems.-} For simplicity, we
take the two parametric driving frequencies satisfying $\omega_{d1}=\omega_{d2}=\omega_{d}$;
then to diagonalize $H_{c}$, we define the squeezing operator $a_{sj}$
via the transformation \cite{Xin2015}:
\begin{align}
a_{j} & =\cosh\left(r_{dj}\right)a_{sj}-e^{-i\Phi_{dj}}\sinh\left(r_{dj}\right)a_{sj}^{\dagger},
\end{align}
where $r_{dj}=\left(1/4\right)\ln\left[\left(\Delta_{j}+2\varLambda_{j}\right)/\left(\Delta_{j}-2\varLambda_{j}\right)\right]$
and $\Delta_{j}=\omega_{j}-\omega_{d}/2$, which requires $\left|\Delta_{j}\right|>\left|2\varLambda_{j}\right|$
to avoid the system instability. Due to the photon-hopping, the transformational
optical modes are coupled though the coherent term $\lambda_{1}a_{s1}a_{s2}^{\dagger}+\mathrm{H.c.}$
and squeezing term $\lambda_{2}a_{s1}a_{s2}+\mathrm{H.c.}$, in which
$\lambda_{1,2}$ is the effective photon-hopping.

For convenient discussion, we define the effective coupling ratio
between the squeezing and coherent terms $f_{1}\equiv\left|\frac{\lambda_{2}\left(\omega_{s1}-\omega_{s2}\right)}{\lambda_{1}\left(\omega_{s1}+\omega_{s2}\right)}\right|$,
where $\omega_{s1,2}$ is the frequency of the transformational optical
mode $a_{s1,2}$. With the rotating wave approximation, it is obvious
that we can reserve the squeezing (coherent) term for $f_{1}\gg1$
($f_{1}\ll1$), which can be used to realize different optomechanical
interactions.

When we have $f_{1}\gg1$, the squeezing term can be used to enhance
optomechanical coupling strength \cite{Li2016}, and we can further
diagonalize the two-mode squeezing terms via the squeezing transformation
($j\neq k$)
\begin{equation}
a_{sj}=\cosh\left(r\right)A_{j}-e^{-i\Phi}\sinh\left(r\right)A_{k}^{\dagger},\label{eq:trans}
\end{equation}
with $A_{j}$ is the annihilation operator for the supermode $j$
with frequency $W_{j\left(k\neq j\right)}=\omega_{sj}\cosh^{2}\left(r\right)+\omega_{sk}\sinh^{2}\left(r\right)-\frac{\left|J'\right|\sinh\left(2r\right)}{2}$.
The effective interaction Hamiltonian can be rewritten as
\begin{eqnarray}
H_{int} & = & -\sum_{j=1}^{2}G_{j}A_{j}^{\dagger}A_{j}\left(b^{\dagger}+b\right)\nonumber \\
 &  & +\sum_{j\leq k=1}^{2}\left(G_{jk}A_{j}A_{k}+\mathrm{H.c.}\right)\left(b^{\dagger}+b\right)\nonumber \\
 &  & -\left(G_{p12}A_{1}^{\dagger}A_{2}+\mathrm{H.c.}\right)\left(b^{\dagger}+b\right),\label{eq:Hint}
\end{eqnarray}
which describes the typical optomechanical forms including the radiation-pressure,
parametric amplification, and three-mode optomechanical couplings.
Here $G_{j}$ is the effective coupling of optomechanical systems,
where
\begin{align}
G_{1} & =g_{0}\cosh\left(2r_{d2}\right)\sinh^{2}\left(r\right),\\
G_{2} & =g_{0}\cosh\left(2r_{d2}\right)\cosh^{2}\left(r\right),
\end{align}
with
\begin{eqnarray}
r & = & \frac{1}{4}\ln\frac{\omega_{s1}+\omega_{s2}+\left|J'\right|}{\omega_{s1}+\omega_{s2}-\left|J'\right|},\\
\frac{J'}{2J} & = & e^{i\Phi_{d2}}\left[\cosh\left(r_{d1}\right)\sinh\left(r_{d2}\right)\right.\nonumber \\
 &  & +\left.\cosh\left(r_{d2}\right)\sinh\left(r_{d1}\right)e^{i\Delta\Phi}\right],
\end{eqnarray}
depending on the phase difference $\Delta\Phi=\Phi_{d1}-\Phi_{d2}$.
As illustrated in Fig. 1(b), the phase difference $\Delta\Phi$ determines
the effective optomechanical couplings. As a comparison with the previous
proposals \cite{Xin2015,Li2016}, the coupling $G_{j}$ is greatly
enhanced as the product of enhancement from the single-mode \cite{Xin2015}
and two-mode \cite{Li2016} squeezing. Here $\Phi=\mathrm{arg}\left(J'\right)$
and the explicit expressions for the parameters $\lambda_{1,2}$,
$\omega_{sj}$, $G_{jk}$, and $G_{p12}$ can be found in the Supplemental
Material \cite{Zhang2017}.

\begin{figure}
\includegraphics[width=0.95\columnwidth]{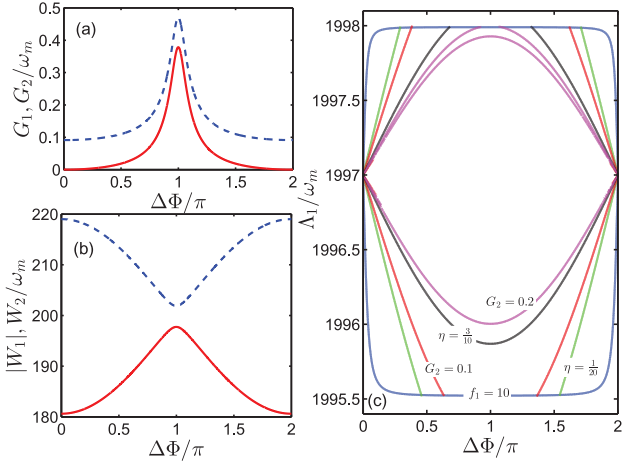}\caption{(Color online) (a) The coupling $G_{1}/\omega_{m}$ (red-solid line)
and $G_{2}/\omega_{m}$ (blue-dashed line) versus phase difference
$\Delta\Phi$. (b) The supermodes $\left|W_{1}\right|/\omega_{m}$
(red-solid line) and $W_{2}/\omega_{m}$ (blue-dashed line) versus
phase difference $\Delta\Phi$. The driving amplitude is $\varLambda_{1}=1997.96\omega_{m}$.
(c) Equipotential lines versus $\varLambda_{1}$ and $\Delta\Phi$.
The other parameters are $\Delta_{1}=-4000\omega_{m}$, $\Delta_{2}=4000\omega_{m}$,
$\varLambda_{2}=1997\omega_{m}$, $g_{0}=0.005\omega_{m}$, $J=0.95\omega_{m}$,
and $\kappa=0.05\omega_{m}$.}

\label{Fig2}
\end{figure}

In Fig.$\,$\ref{Fig2}(a), the optomechanical coupling strengths
$G_{1,2}$ are plotted with reasonable parameters, which demonstrate
the significant enhancement by controlling the phase $\Delta\Phi$
and show the strong-coupling regime is achievable (i.e. $G_{1},~G_{2}\sim\omega_{m}>\kappa$)
for $\Delta\Phi$ around the optimal $\Delta\Phi=\pi$. Because we
choose the parametric pump detuning $\Delta_{1}<0$ and $\Delta_{2}>0$
, which lead to $r_{d1}<0$ and $r_{d2}>0$, the effective $\left|J'\right|$
reaches its maximum when $\Delta\Phi=\pi$ and the minimum when $\Delta\Phi=0$.
In the Fig.$\,$\ref{Fig2}(b), we plot the dependence of supermode
frequencies $\left|W_{1}\right|/\omega_{m}$, $W_{2}/\omega_{m}$
on $\Delta\Phi$. When $\Delta\Phi$ tends to $0$ or $2\pi$, we
have the coupling strength $G_{2}\gg G_{1}$, while the other couplings
$G_{jk}$ and $G_{p12}$ can be ignored for $\left|W_{j}+W_{k}\pm\omega_{m}\right|\gg G_{jk}$
and $\left|W_{1}-W_{2}\pm\omega_{m}\right|\gg G_{p12}$. To show the
enhanced coupling strengths for different driving amplitude $\varLambda_{1}$
and phase $\Delta\Phi$, we plot the equipotential lines of $f_{1}$,
$G_{1}/\omega_{m}$, $G_{2}/\omega_{m}$, and $\eta=G_{1}/G_{2}$
in the Fig.$\,$\ref{Fig2}(c). The inner region surrounded by the
blue line $f_{1}=10\gg1$ means that only the squeezing term dominates
and the rotating wave approximation is appropriate. The red line $G_{2}=0.1\omega_{m}>\kappa$
and green line $\eta=\frac{1}{20}$ show that only the second optomechanical
coupling reaches the strong-coupling regime. When the parameters $\varLambda_{1}$
and $\Delta\Phi$ tend to the central area, as shown by both the pink
and black lines, both $G_{1}$ and $G_{2}$ can reach strong-coupling
regime.

With appropriate parameters, the parametric amplification coupling
forms in Eq.$\,$\ref{eq:Hint} can also been obtained when $\omega_{m}\gg G_{j}$
and $\left|W_{1}-W_{2}\pm\omega_{m}\right|\gg G_{p12}$, and meanwhile
the frequency matching $\left|W_{j}+W_{k}\pm\omega_{m}\right|\approx0$
is satisfied. The detailed discussion for the parametric amplification
can be found in the Supplemental Material \cite{Zhang2017}. Compared
to previous schemes that also employ the parametric interaction \cite{Liu2013,Lemonde2013,B=0000F8rkje2013},
the coefficient of parametric amplification is further improved by
our coupled-cavity configuration, which can be used to generate photon-phonon
pairs. Even when only one parametric driving field exists in the cavity,
the optomechanical coupling can still be enhanced than no parametric
driving \cite{Zhang2017}.

\emph{Phase-controlled phonon laser.-} The laser term in Eq.$\,$\ref{eq:Hint}
could be utilized for realizing the phonon laser if $G_{p12}$ is
dominated over other coupling strengths. This interaction is a triply-resonant
interaction, with the advantage that the pump and idle optical field
are resonantly enhanced. When the triply-resonant frequency $W_{1}-W_{2}\approx\omega_{m}$
is matched, we have the parameter $f_{1}\ll1$. By a similar transformation
\cite{Zhang2017} with the Eq.$\,$\ref{eq:trans}, we obtain

\begin{align}
G_{p12} & =-\frac{g_{0}}{2}e^{-i\Phi}\cosh\left(2r_{d2}\right)\sin\left(\theta\right),
\end{align}
with $\Phi=\mathrm{arg}\left(J'\right)$ and $\theta=\arctan\left[\left|J'\right|/\left(\omega_{s2}-\omega_{s1}\right)\right]$,
in which $J'=2J\left[\cosh\left(r_{d1}\right)\cosh\left(r_{d2}\right)+\sinh\left(r_{d1}\right)\sinh\left(r_{d2}\right)e^{i\Delta\Phi}\right]$.

\begin{figure}
\includegraphics[width=1\columnwidth]{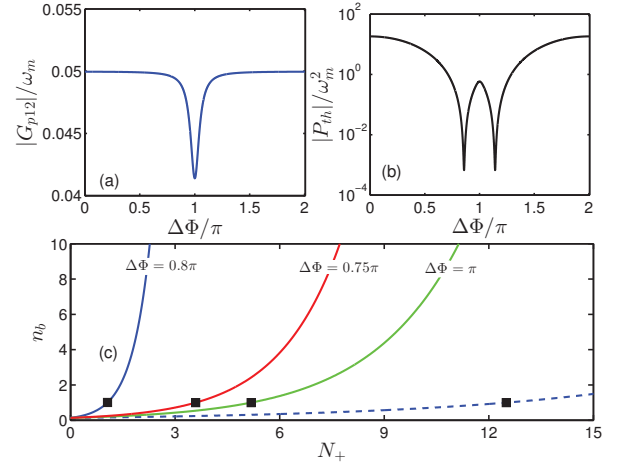}\caption{(Color online) (a) The coupling $\left|G_{p12}\right|/\omega_{m}$
versus phase difference $\Delta\Phi$. (b) The threshold pump power
$P_{th}/\omega_{m}^{2}$ versus difference phase $\Delta\Phi$. (c)
Plot of the stimulated emitted phonon number $n_{b}$ as a function
of the density of the supermode $A_{1}$. The threshold density $N_{+}$
denoted by the black square points is obtained for $\mathscr{G}=\gamma_{m}$.
The parameters are $\Delta_{1}=20\omega_{m}$, $\Delta_{2}=100\omega_{m}$,
$\varLambda_{1}=9.94\omega_{m}$, $\varLambda_{2}=49.99\omega_{m}$,
$J=0.1\omega_{m}$, $g_{0}=0.002\omega_{m}$, $\kappa=0.05\omega_{m}$,
and $\gamma_{m}=0.001\omega_{m}$. }

\label{Fig3}
\end{figure}

In Fig.$\,$\ref{Fig3}(a), we plot the triply-resonant phonon lasing
coupling strength $\left|G_{p12}\right|$ versus the phase difference
$\Delta\Phi$, which can reach the strong-coupling regime $\left|G_{p12}\right|\simeq\kappa$,
and there is no obvious change with the increasing of the phase difference.
When the frequency matches $W_{1}-W_{2}\approx\omega_{m}$, we have
$\left|G_{j}\right|/\omega_{m},~\left|G_{jk}/\left(W_{j}+W_{k}\pm\omega_{m}\right)\right|\ll1$,
therefore, the other coupling strengthscan be neglected.

If the effective optical cavity decay rate exceeds the mechanical
dissipation rate ($\kappa\gg\gamma_{m}$), we find the mechanical
gain \cite{Grudinin2010}
\begin{equation}
\mathscr{G}=\frac{\left|G_{p12}\right|^{2}\Delta N\kappa}{\left(W_{1}-W_{2}-\omega_{m}\right)^{2}+\left(\kappa/2\right)^{2}},
\end{equation}
where $\Delta N=N_{+}-N_{-}\approx N_{+}$ with $N_{+}=A_{1}^{\dagger}A_{1}$
and $N_{-}=A_{2}^{\dagger}A_{2}$. The gain has a spectral bandwidth
$\kappa$ and $W_{1}-W_{2}=\omega_{m}$ is corresponding to the maximum
gain.

The threshold condition $\mathscr{G}=\gamma_{m}$ determines the emitted
phonon number, which is shown in Fig.$\,$\ref{Fig3}(c). The solid
lines are stimulated emitted phonon number $n_{b}\left[\gamma_{m}\right]=\exp\left[2\left(\mathscr{G}-\gamma_{m}\right)/\gamma_{m}\right]$
as a function of the density $N_{+}$ for different $\Delta\Phi$.
If there is no any OPA in the cavity, the emitted phonon number $n_{b}$
with the resonance $W_{1}-W_{2}=\omega_{m}$ is shown by the dashed
line in Fig.$\,$\ref{Fig3}(c). Clearly, it indicates an ultralow-threshold
phonon laser by tuning the phase difference $\Delta\Phi$.

The black square points denote the threshold density $N_{+}$ for
$\mathscr{G}=\gamma_{m}$ in Fig.$\,$\ref{Fig3}(c). We know the
threshold pump power as $P_{th}=N_{+}\kappa W_{1}$, and we obtain
\begin{equation}
P_{th}\approx\frac{\gamma_{m}W_{1}\left[\left(W_{1}-W_{2}-\omega_{m}\right)^{2}+\left(\kappa/2\right)^{2}\right]}{\left|G_{p12}\right|^{2}},
\end{equation}
which is plotted as the function of phase difference $\Delta\Phi$
in Fig.$\,$\ref{Fig3}(b). There are two dips, which mean an ultralow-threshold
power with the near resonance $W_{1}-W_{2}\approx\omega_{m}$. The
ultralow-threshold power $P_{th}$ is related to the frequency difference
$W_{1}-W_{2}$ controlled by the strengths and phases of parametric
driving terms. From the Fig. 3, it is noted that the threshold density
$N_{+}\leq1$ can be obtained by changing the phase difference $\Delta\Phi$.
In other words, the phonon lasing is possible with an ultralow-threshold
power, as low as single photon.

\emph{Discussion.- }In the presence of a parametric drive, the noise
from the optical cavity decay might also be amplified. To circumvent
the amplified noise, a possible strategy is to introduce a broadband
single-mode or two-mode squeezed vacuum via dissipative squeezing
\cite{Xin2015,Clerk2016,Li2016}. This steady-state technique has
recently been implemented experimentally \cite{Wollman2015,Pirkk2015,Lecocq2015},
and recently it has been experimentally demonstrated that squeezed
light can be used to cool the motion of a macroscopic mechanical object
without resolved-sideband condition \cite{Clark2016}. One can also
take advantage of the tunability of the parametric drive to avoid
significant perturbation of the initial photon state \cite{Clerk2016}.
It is feasible to suppress the cavity noise in the experiment for
realizing the optomechanical strong-coupling regime.

\emph{Conclusion.-} we present a scheme for enhancing phase-controlled
optomechanical couplings into the single-photon strong-coupling regime
by optical squeezing. With two OPAs in two coupled optical cavities,
we obtain the squeezing of transformational optical modes, which leads
to exponentially enhanced optomechanical systems. The phase difference
between the two driving fields on OPAs can control the enhanced radiation-pressure,
parametric amplification, and three-mode optomechanical couplings.
In particular, the three-mode optomechanical coupling can be used
to realize a low-threshold phonon laser, and the threshold pump power
is decreased greatly with the giant enhancement of mechanical gain.
With current experimentally accessible parameters, our scheme should
be feasible to study quantum optomechanics. This allows us to explore
a number of interesting quantum optomechanics applications ranging
from single-photon sources to nonclassical quantum states.

{\em Acknowledgments.} We thank Yue-Man Kuang, Xin-You L\"{u},
Ya-Feng Jiao, and Jie-Qiao Liao for useful suggestions. This work
was funded by the National Key R \& D Program (Grants No. 2016YFA0301300
and No. 2016YFA0301700), the National Natural Science Foundation of
China (Grants No. 11474271, No. 11674305, and No. 61505195), and the
China Postdoctoral Science Foundation (No. 2016M602013). H. J. is
supported by the National Natural Science Foundation of China (Grants
No. 11474087 and No. 11422437).

\newpage
\onecolumngrid

\setcounter{section}{0} \setcounter{equation}{0} \setcounter{figure}{0} \setcounter{table}{0} %
\renewcommand{\thesection}{S-\arabic{section}}
\renewcommand{\theequation}{S\arabic{equation}}
\renewcommand{\thefigure}{S\arabic{figure}}
\renewcommand{\bibnumfmt}[1]{[S#1]}
\renewcommand{\citenumfont}[1]{S#1}

\newpage

\begin{center}\large
\textbf{Supplementary Materials}
\end{center}

\section{effective Hamiltonian}

From the main text, we know that the Hamiltonian of the system can
be written as
\begin{align}
H= & H_{c}+H_{m}+J\left(a_{1}a_{2}^{\dagger}+a_{1}^{\dagger}a_{2}\right).
\end{align}

For simplicity, we take the two parametric driving frequencies satisfying
$\omega_{d1}=\omega_{d2}=\omega_{d}$. In the interaction picture
$H_{0}=\frac{\omega_{d}}{2}\left(a_{1}^{\dagger}a_{1}+a_{2}^{\dagger}a_{2}\right)$,
the Hamiltonian of the system can be written as
\begin{align}
H= & \sum_{j=1}^{2}\Delta_{j}a_{j}^{\dagger}a_{j}+\varLambda_{j}\left(a_{j}^{\dagger2}e^{-i\Phi_{dj}}+\mathrm{H.c.}\right)+\omega_{m}b^{\dagger}b-g_{0}a_{2}^{\dagger}a_{2}\left(b^{\dagger}+b\right)+J\left(a_{1}a_{2}^{\dagger}+a_{1}^{\dagger}a_{2}\right),
\end{align}
where the detuning $\Delta_{j}=\omega_{j}-\omega_{d}/2$.

To diagonalize the $H_{c}$, we introduce a squeezing transformation
\cite{Xin2015}
\begin{equation}
a_{j}=\cosh\left(r_{dj}\right)a_{sj}-e^{-i\Phi_{dj}}\sinh\left(r_{dj}\right)a_{sj}^{\dagger},
\end{equation}
where
\begin{equation}
r_{dj}=1/4\ln\left[\left(\Delta_{j}+2\varLambda_{j}\right)/\left(\Delta_{j}-2\varLambda_{j}\right)\right],
\end{equation}
which requires $\left|\Delta_{j}\right|>\left|2\varLambda_{j}\right|$
  to avoid the system instable.

The Hamiltonian of the system can be changed into
\begin{eqnarray}
H & = & \sum_{j=1}^{2}\omega_{sj}a_{sj}^{\dagger}a_{sj}+\omega_{m}b^{\dagger}b-g_{s2}a_{s2}^{\dagger}a_{s2}\left(b^{\dagger}+b\right)+g_{p2}\left(e^{-i\Phi_{d2}}a_{s2}^{\dagger2}+\mathrm{H.c.}\right)\left(b^{\dagger}+b\right)\nonumber \\
 &  & +J\left(\lambda_{1}a_{s1}a_{s2}^{\dagger}-\lambda_{2}a_{s1}a_{s2}+\mathrm{H.c.}\right)-F\left(b^{\dagger}+b\right)+C
\end{eqnarray}
where
\begin{align}
\omega_{sj} & =\left(\Delta_{j}-2\varLambda_{j}\right)\exp\left(2r_{dj}\right),\\
g_{s2} & =g_{0}\left[\sinh^{2}\left(r_{d2}\right)+\cosh^{2}\left(r_{d2}\right)\right]=\frac{g_{0}\Delta_{2}}{\sqrt{\Delta_{2}^{2}-4\varLambda_{2}^{2}}},\\
g_{p2} & =g_{0}\cosh\left(r_{d2}\right)\sinh\left(r_{d2}\right)=\frac{g_{0}\varLambda_{2}}{\sqrt{\Delta_{2}^{2}-4\varLambda_{2}^{2}}},\\
\lambda_{1} & =\cosh\left(r_{d1}\right)\cosh\left(r_{d2}\right)+\sinh\left(r_{d1}\right)\sinh\left(r_{d2}\right)e^{i\left(\Phi_{d1}-\Phi_{d2}\right)},\\
\lambda_{2} & =\cosh\left(r_{d1}\right)\sinh\left(r_{d2}\right)e^{i\Phi_{d2}}+\sinh\left(r_{d1}\right)\cosh\left(r_{d2}\right)e^{i\Phi_{d1}},\\
F & =g_{0}\sinh^{2}\left(r_{d2}\right),\\
C & =\sum_{j=1}^{2}\Delta_{j}\sinh^{2}\left(r_{dj}\right)-2\varLambda_{j}\cosh\left(r_{dj}\right)\sinh\left(r_{dj}\right).
\end{align}

\section{phase-controlled optomechanical systems with $f_{1}\gg1$}

With the rotating approximation, we can eliminate the term $\lambda_{1}a_{s1}a_{s2}^{\dagger}+\mathrm{H.c.}$
when we have
\begin{equation}
f_{1}\equiv\left|\frac{\lambda_{2}\left(\omega_{s1}-\omega_{s2}\right)}{\lambda_{1}\left(\omega_{s1}+\omega_{s2}\right)}\right|\gg1,
\end{equation}
which means that the effective interaction from squeezing terms is
much larger. It is obvious that only the squeezing terms can be reserved
to ehance optomechanical coupling strength \cite{Li2016}.

Similar to the above, we can diagonalize the two-mode squeezing via
the squeezing transformation \cite{Li2016}
\begin{equation}
a_{sj}=\cosh\left(r\right)A_{j}-e^{-i\Phi}\sinh\left(r\right)A_{k}^{\dagger}\left(j\neq k\right),
\end{equation}
where
\begin{equation}
r=\frac{1}{4}\ln\left[\left(\omega_{s1}+\omega_{s2}+\left|J'\right|\right)/\left(\omega_{s1}+\omega_{s2}-\left|J'\right|\right)\right],
\end{equation}
in which $J'=2J\lambda_{2}$ and $\Phi=\mathrm{arg}\left(J'\right)$.
To avoid the system instability, we need
\begin{equation}
f_{2}\equiv\left|\omega_{s1}+\omega_{s2}\right|-\left|J'\right|>0.
\end{equation}
This leads to the following Hamiltonian
\begin{eqnarray}
H & = & \omega_{m}b^{\dagger}b+\sum_{j=1}^{2}W_{j}A_{j}^{\dagger}A_{j}-G_{j}A_{j}^{\dagger}A_{j}\left(b^{\dagger}+b\right)+\sum_{j\leq k=1}^{2}\left(G_{jk}A_{j}A_{k}+\mathrm{H.c.}\right)\left(b^{\dagger}+b\right)\nonumber \\
 &  & -\left(G_{p12}A_{1}^{\dagger}A_{2}+\mathrm{H.c.}\right)\left(b^{\dagger}+b\right)+\left(F'-F\right)\left(b^{\dagger}+b\right)+C+C^{'},
\end{eqnarray}
where
\begin{align}
W_{1} & =\omega_{s1}\cosh^{2}\left(r\right)+\omega_{s2}\sinh^{2}\left(r\right)-\left|J'\right|\sinh\left(2r\right)/2,\\
W_{2} & =\omega_{s2}\cosh^{2}\left(r\right)+\omega_{s1}\sinh^{2}\left(r\right)-\left|J'\right|\sinh\left(2r\right)/2,\\
G_{1} & =g_{0}\cosh\left(2r_{d2}\right)\sinh^{2}\left(r\right),\\
G_{2} & =g_{0}\cosh\left(2r_{d2}\right)\cosh^{2}\left(r\right),\\
G_{12} & =\frac{g_{0}}{2}e^{i\Phi}\cosh\left(2r_{d2}\right)\sinh\left(2r\right),\\
G_{11} & =\frac{g_{0}}{2}e^{i\left(2\Phi-\Phi_{d2}\right)}\sinh\left(2r_{d2}\right)\sinh^{2}\left(r\right),\\
G_{22} & =\frac{g_{0}}{2}e^{i\Phi_{d2}}\sinh\left(2r_{d2}\right)\cosh^{2}\left(r\right),\\
G_{p12} & =\frac{g_{0}}{2}e^{i\left(\Phi_{d2}-\Phi\right)}\sinh\left(2r_{d2}\right)\sinh\left(2r\right),\\
F' & =g_{0}\cosh\left(2r_{d2}\right)\sinh^{2}\left(r\right),\\
C^{'} & =\left(\omega_{s2}+\omega_{s1}\right)\sinh^{2}\left(r\right)-\left|J'\right|\sinh\left(r\right)\cosh\left(r\right).
\end{align}
We notice that $F'-F$ and $C+C'$ are only the displaced term and
a constant, respectively, which can be neglected in the optomechanical
system.

We have discussed the radiation-pressure optomechanical coupling in
the main text. With the appropriate parameters, the parametric amplification
coupling forms can also been obtained.

\begin{figure}[tb]
\includegraphics[width=0.9\columnwidth]{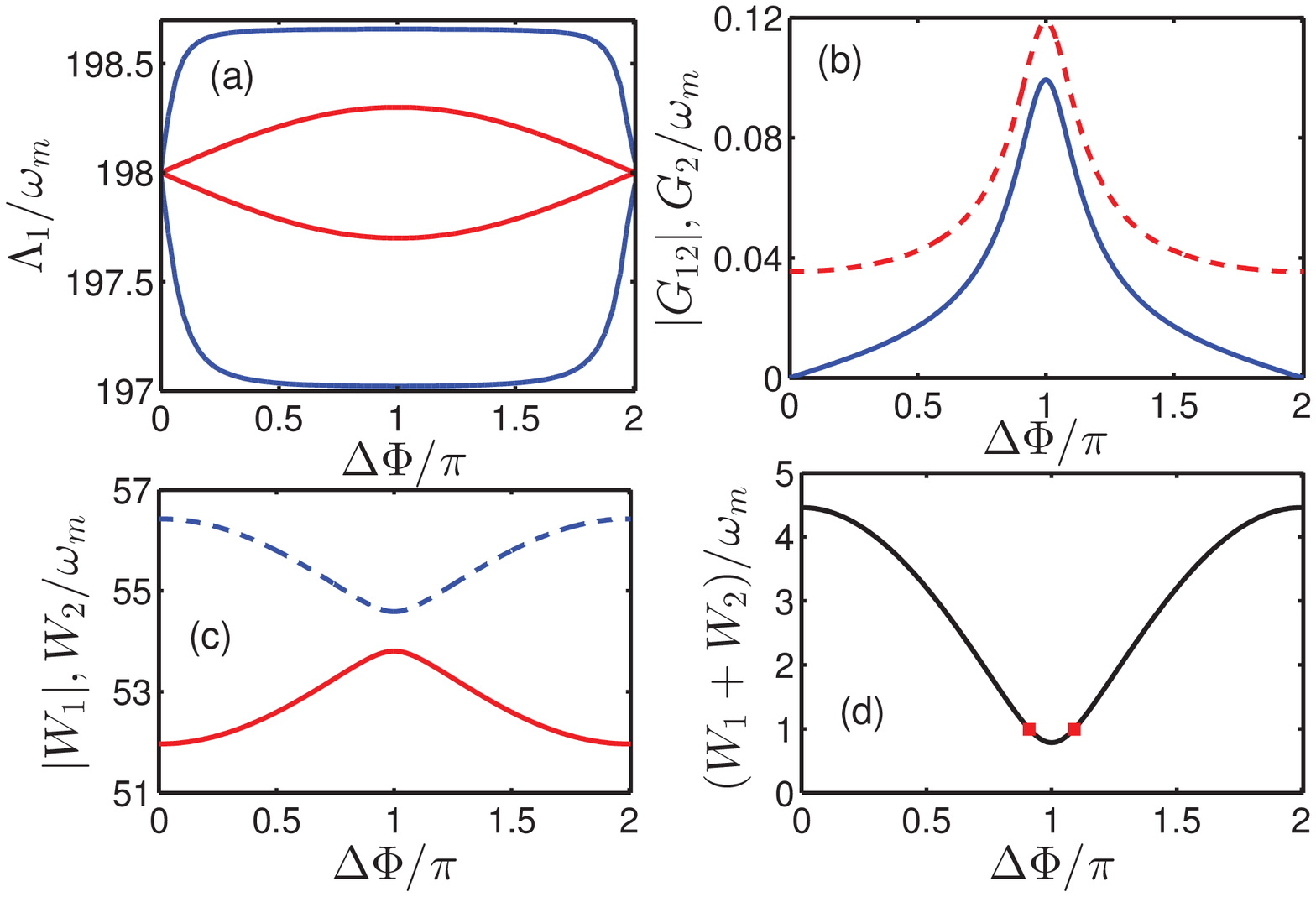}

\caption{(Color online) (a) Equipotential lines $f_{1}=10$ (blue line) and
$f_{2}=0$ (red line) versus $\varLambda_{1}$ and $\Delta\Phi$.
(b) The coupling $G_{2}/\omega_{m}$ (red-dashed line) and $G_{12}/\omega_{m}$
(blue-solid line) versus phase difference $\Delta\Phi$. (c) The supermodes
$\left|W_{1}\right|/\omega_{m}$ (red-solid line) and $W_{2}/\omega_{m}$
(blue-dashed line) versus phase difference $\Delta\Phi$. (d)The frequency
$\left(W_{1}+W_{2}\right)/\omega_{m}$ versus phase difference $\Delta\Phi$.
The red square points show the resonant condition $W_{1}+W_{2}=\omega_{m}$.
The parameters are $\Delta_{1}=-400\omega_{m}$, $\Delta_{2}=400\omega_{m}$,
$\varLambda_{1}=198.305\omega_{m}$, $\varLambda_{2}=198\omega_{m}$,
$g_{0}=0.005\omega_{m}$, $J=0.3\omega_{m}$, and $\kappa=0.05\omega_{m}$.}

\label{FigS1}
\end{figure}

To obtain the effective Hamiltonian , we notice that there are the
followling conditions: (a) $f_{1}\gg1$ (rotating wave approximation);
(b) $\left|\Delta_{j}\right|>\left|2\varLambda_{j}\right|$ and $f_{2}>0$
(stable conditions). Naturally, the system parameters are chosen to
satisfy $\left|\Delta_{j}\right|>\left|2\varLambda_{j}\right|$. Equipotential
lines $f_{1}=10$ (blue line) and $f_{2}=0$ (red line) versus $\varLambda_{1}$
and $\Delta\Phi$ are plotted in Fig.$\,$\ref{FigS1}(a), and the
area between the blue and red lines fully satisfies the above conditions.
In Fig.$\,$\ref{FigS1}(b), we plot the coupling $G_{2}/\omega_{m}$
(red-dashed line) and $G_{12}/\omega_{m}$ (blue-solid line) versus
phase difference $\Delta\Phi$, and we can reach the strong-coupling
regime when $\Delta\Phi=\pi$, however, which is much smaller than
the mechanical frequency $\omega_{m}$. The supermode frequencies
$\left|W_{1}\right|/\omega_{m}$ (red-solid line) and $W_{2}/\omega_{m}$
(blue-dashed line) are shown in Fig.$\,$\ref{FigS1}(c), which means
that we have $\omega_{m}\gg G_{j}$, $\left|2W_{j}\pm\omega_{m}\right|\gg\left|G_{jj}\right|$
and $\left|W_{1}-W_{2}\pm\omega_{m}\right|\gg\left|G_{p12}\right|$.
While we find the frequency matching (red square points) $\left|W_{1}+W_{2}-\omega_{m}\right|\approx0$
inFig.$\,$\ref{FigS1}(d), which leads that only the term $G_{12}$
can be reserved. It is obvious that we can also obtain the parametric
amplification coupling forms when $\left|2W_{j}\pm\omega_{m}\right|\approx0$
with appropriate parameters.

When only one parametric driving field exists in the cavity, we can
still realize enhanced optomechanical coupling without phase control.
If the parametric driving field exists in the second cavity, it means
$\Lambda_{1}=0$ $\left(r_{d1}=0\right)$, and all coupling forms
are same to the above. The phase difference does not appear in the
expression of effective coupling $\left|J'\right|=2J\sinh\left(r_{d2}\right)$,
which means the coupling parameters can not be tuned by the phase
difference. It needs a stronger photon-hopping interaction $J$ because
of no product factor $\sinh\left(r_{d1}\right)$ or $\cosh\left(r_{d1}\right)$
in the effective $\left|J'\right|$. We can still realize the controlled
optomechanical coupling by tuning $\Lambda_{2}$, $\Delta_{j}$ and
$J$. If the parametric driving field exists in the first cavity,
we have $\Lambda_{2}=0$ $\left(r_{d2}=0\right)$. The coupling strength
becomes $G_{1}=g_{0}\sinh^{2}\left(r\right)$ and $G_{2}=g_{0}\cosh^{2}\left(r\right)$,
and the enhanced optomechanical coupling can still be obtained. The
OPA is put into the auxiliary cavity, which may be easier to implement
in the experiment.

\section{phase-controlled phonon laser with $f_{1}\ll1$}

When $f_{1}\ll1$, we can neglect the term $\lambda_{2}a_{s1}^{\dagger}a_{s2}^{\dagger}+\mathrm{H.c.}$
(rotating wave approximation), and the Hamiltonian of the system can
be written as
\begin{eqnarray}
H & = & \sum_{j=1}^{2}\omega_{sj}a_{sj}^{\dagger}a_{sj}+\omega_{m}b^{\dagger}b-g_{s2}a_{s2}^{\dagger}a_{s2}\left(b^{\dagger}+b\right)+g_{p2}\left(e^{-i\Phi_{d2}}a_{s2}^{\dagger2}+\mathrm{H.c.}\right)\left(b^{\dagger}+b\right)\nonumber \\
 &  & +J\left(\lambda_{1}a_{s1}a_{s2}^{\dagger}+\mathrm{H.c.}\right)-F\left(b^{\dagger}+b\right)+C,
\end{eqnarray}
To diagonalize the interaction term $\lambda_{1}$, we introduce the
transformation
\begin{align}
a_{s1} & =\cos\left(\frac{\theta}{2}\right)A_{1}+e^{-i\Phi}\sin\left(\frac{\theta}{2}\right)A_{2},\\
a_{s2} & =\cos\left(\frac{\theta}{2}\right)A_{2}-e^{i\Phi}\sin\left(\frac{\theta}{2}\right)A_{1},
\end{align}
where $\theta=\arctan\left(\frac{\left|J'\right|}{\omega_{s2}-\omega_{s1}}\right)$,
in which $J'=2J\lambda_{1}$ and $\Phi=arg\left(J'\right)$.

The Hamiltonian can be written as
\begin{eqnarray}
H & = & \omega_{m}b^{\dagger}b+\sum_{j=1}^{2}W_{j}A_{j}^{\dagger}A_{j}-G_{j}A_{j}^{\dagger}A_{j}\left(b^{\dagger}+b\right)+\sum_{j\leq k=1}^{2}\left(G_{jk}A_{j}A_{k}+\mathrm{H.c.}\right)\left(b^{\dagger}+b\right)\nonumber \\
 &  & +\left(G_{p12}A_{1}^{\dagger}A_{2}+\mathrm{H.c.}\right)\left(b^{\dagger}+b\right)+-F\left(b^{\dagger}+b\right)+C,
\end{eqnarray}
where
\begin{align}
W_{1} & =\omega_{s1}\cos^{2}\left(\frac{\theta}{2}\right)+\omega_{s2}\sin^{2}\left(\frac{\theta}{2}\right)-\left|J'\right|\sin\left(\theta\right)/2,\\
W_{2} & =\omega_{s2}\cos^{2}\left(\frac{\theta}{2}\right)+\omega_{s1}\sin^{2}\left(\frac{\theta}{2}\right)+\left|J'\right|\sin\left(\theta\right)/2,\\
G_{1} & =g_{0}\cosh\left(2r_{d2}\right)\sin^{2}\left(\frac{\theta}{2}\right),\\
G_{2} & =g_{0}\cosh\left(2r_{d2}\right)\cos^{2}\left(\frac{\theta}{2}\right),\\
G_{12} & =-\frac{g_{0}}{2}e^{i\left(\Phi_{d2}+\Phi\right)}\sinh\left(2r_{d2}\right)\sin\left(\theta\right),\\
G_{11} & =\frac{g_{0}}{2}e^{i\left(2\Phi+\Phi_{d2}\right)}\sinh\left(2r_{d2}\right)\sin^{2}\left(\frac{\theta}{2}\right),\\
G_{22} & =\frac{g_{0}}{2}e^{i\Phi_{d2}}\sinh\left(2r_{d2}\right)\cos^{2}\left(\frac{\theta}{2}\right),\\
G_{p12} & =\frac{g_{0}}{2}e^{-i\Phi}\cosh\left(2r_{d2}\right)\sin\left(\theta\right).
\end{align}
We know that the phonon laser can be realized by the Hamiltonian
\begin{eqnarray}
H & = & \omega_{m}b^{\dagger}b+\sum_{j=1}^{2}W_{j}A_{j}^{\dagger}A_{j}+G_{p12}A_{1}^{\dagger}A_{2}b+\mathrm{H.c.},
\end{eqnarray}
when we have $\omega_{m}\gg G_{j}$, $\left|W_{j}+W_{k}\pm\omega_{m}\right|\gg\left|G_{jk}\right|$
and $\left|W_{1}-W_{2}\pm\omega_{m}\right|\approx0$.

\begin{figure}[tb]
\includegraphics[width=0.9\columnwidth]{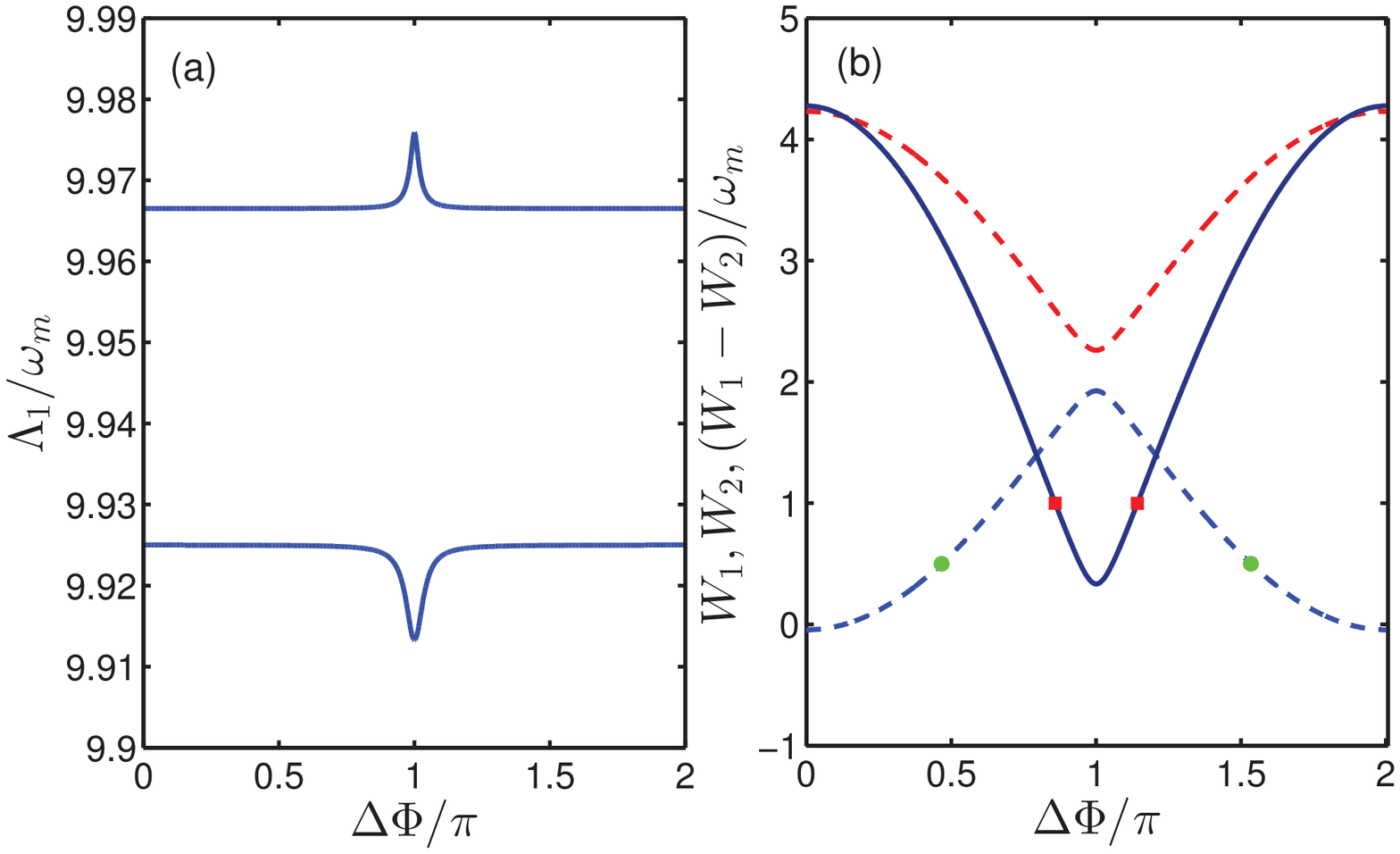}

\caption{(Color online) (a) Equipotential lines $f_{1}=0.1$ versus $\varLambda_{1}$
and $\Delta\Phi$. (b) The coupling $G_{2}/\omega_{m}$ (red-dashed
line) and $G_{12}/\omega_{m}$ (blue-solid line) versus phase difference
$\Delta\Phi$. (c) The frequencies $W_{1}/\omega_{m}$ (red-dashed
line) , $W_{2}/\omega_{m}$ (blue-dashed line), and $\left(W_{1}-W_{2}\right)/\omega_{m}$
versus phase difference $\Delta\Phi$. The red square points and green
dots show the resonant condition $W_{1}-W_{2}=\omega_{m}$ and $W_{2}=0.5\omega_{m}$,
respectively. The other parameters are $\Delta_{1}=20\omega_{m}$,
$\Delta_{2}=100\omega_{m}$, $\varLambda_{1}=9.94\omega_{m}$, $\varLambda_{2}=49.99\omega_{m}$,
$g_{0}=0.002\omega_{m}$, $J=0.1\omega_{m}$, and $\kappa=0.05\omega_{m}$.}

\label{FigS2}
\end{figure}

In Fig.$\,$\ref{FigS2}(a), we plot the equipotential lines $f_{1}=0.1$
(blue line) versus $\varLambda_{1}$ and $\Delta\Phi$, and the area
between the blue and red lines fully satisfies $f_{1}\ll1$. The frequencies
$W_{1}/\omega_{m}$ (red-dashed line), $W_{2}/\omega_{m}$ (blue-dashed
line), and $\left(W_{1}-W_{2}\right)/\omega_{m}$ (blue-solid line)
are shown in Fig.$\,$\ref{FigS2}(b). We find the frequency matching
(red square points) $\left|W_{1}-W_{2}-\omega_{m}\right|\approx0$,
which leads that only the term $G_{p12}$ can be reserved. While the
rotating wave approximation is satisfied, we can realize the phase-controlled
phonon laser. In Fig.$\,$\ref{FigS2}(b), the frequency $\left|2W_{2}-\omega_{m}\right|\approx0$
can also be matched, which is denoted by the green dots. It is obvious
that we can obtain the parametric amplification coupling form $G_{22}$
only by the tuning phase difference.

\end{document}